# UNCERTAIN CLASSIFICATION OF VARIABLE STARS: HANDLING OBSERVATIONAL GAPS AND NOISE

Nicolás Castro [1], Pavlos Protopapas [2,3], Karim Pichara [1,4,3]

[1]Computer Science Department, Pontificia Universidad Católica de Chile, Santiago, Chile
[2]Harvard-Smithsonian Center for Astrophysics, Cambridge, MA, USA
[3]Institute for Applied Computational Science, Harvard University, Cambridge, MA, USA and
[4]Millenium Institute of Astrophysics, Santiago, Chile
*Draft version January 29, 2018*

### ABSTRACT

Automatic classification methods applied to sky surveys have revolutionized the astronomical target selection process. Most surveys generate a vast amount of time series, or "lightcurves", that represent the brightness variability of stellar objects in time. Unfortunately, lightcurves' observations take several years to be completed, producing truncated time series that generally remain without the application of automatic classifiers until they are finished. This happens because state of the art methods rely on a variety of statistical descriptors or features that present an increasing degree of dispersion when the number of observations decreases, which reduces their precision. In this paper we propose a novel method that increases the performance of automatic classifiers of variable stars by incorporating the deviations that scarcity of observations produces. Our method uses Gaussian Process Regression to form a probabilistic model of each lightcurve's observations. Then, based on this model, bootstrapped samples of the time series features are generated. Finally a bagging approach is used to improve the overall performance of the classification. We perform tests on the MACHO and OGLE catalogs, results show that our method classifies effectively some variability classes using a small fraction of the original observations. For example, we found that RR Lyrae stars can be classified with around 80% of accuracy just by observing the first 5% of the whole lightcurves' observations in MACHO and OGLE catalogs. We believe these results prove that, when studying lightcurves, it is important to consider the features' error and how the measurement process impacts it.

*Subject headings:* variable – stars – machine – learning – astro statistics – data – science – automatic – classification

## 1. INTRODUCTION

Modern synoptic surveys observe giant portions of the sky for long periods of time. This gives astronomers the opportunity to make more and greater discoveries than ever before. Due to its enormity, the generated data can no longer be analyzed by human intensive methods, and the necessity for automatic computational intelligent tools has become unavoidable. In recent years, automatic classification of variable stars through lightcurve analysis has been heavily studied (Debosscher et al. 2007; Wachman et al. 2009; Kim et al. 2009; Wang et al. 2010; Richards et al. 2011; Bloom & Richards 2011; Kim et al. 2011a; Pichara et al. 2012; Bloom et al. 2012; Pichara & Protopapas 2013; Kim et al. 2014; Nun et al. 2014; Mackenzie et al. 2016; Pichara et al. 2016). This task aims to identify certain specific and valuable types of stars, so later they can be studied with greater details by astronomers.

In this line, machine learning techniques have proved to be particularly effective, due to their precision and speed (Debosscher et al. 2007). This kind of tools, train classification models over a group of labeled objects, for example, a significant group of stars whose specific variability type has been previously determined trough spectroscopy. The training process seeks to teach models to recognize underlying patterns that allow them to separate among a set of variability classes. These patterns can be very complex and high dimensional. Fortunately, Machine learning approaches have shown capabilities to discover very complex underlying patterns, that are im-perceptible for human beings (Jiawei & Kamber 2001).

For the task of automatic classification, lightcurves are represented as a vector of statistical features that describe different aspects of them, like brightness variability, color, periodicity, and auto-correlation, among others (Richards et al. 2011; Pichara et al. 2012; Nun et al. 2015) . However, the value of those features is highly dependent on the quality of the measurements on which they are calculated (Kirk & Stumpf 2009). Inherent errors in the values of photometric time-series, as well as the amount of observations, may affect the values of their descriptors. Therefore, the errors committed by classifiers in their predictions can be attributed, at least in part, to the lack of precision of the features used to represent them. In our experiments: we assume that 100% of the observations correspond to the number of points included on each lightcurve after the survey finishes its operation. We also assume that each survey was designed with a specific purpose and that the number of observation was decided according to it. Having that said, any lightcurve with removed observations will be considered as incomplete. If the number of removed observations is considerable we expect to see variations in the values used to represent them. For example, an insufficient amount of observations may result in wrong estimation of periods, in spurious auto-correlations values, or in poorly calculated variability patterns.

Figure 1 shows the value of three different features, for two different lightcurves, calculated at different moments of the observation process. The values of each indi-vidual feature has been normalized and centered around zero, in order to make the variations comparable. It is not surprising that the values change considerably as the number of observations increases, but it is worth noting that stronger changes occur at the beginning, when the number of observations is smaller. This holds for most of the statistical estimates. What is particularly interesting is that this effect is not consistent for different features and for different lightcurves. In fact, it is easy to find cases where the same feature takes longer to stabilize than others or even cases where features do not appear to converge at all.

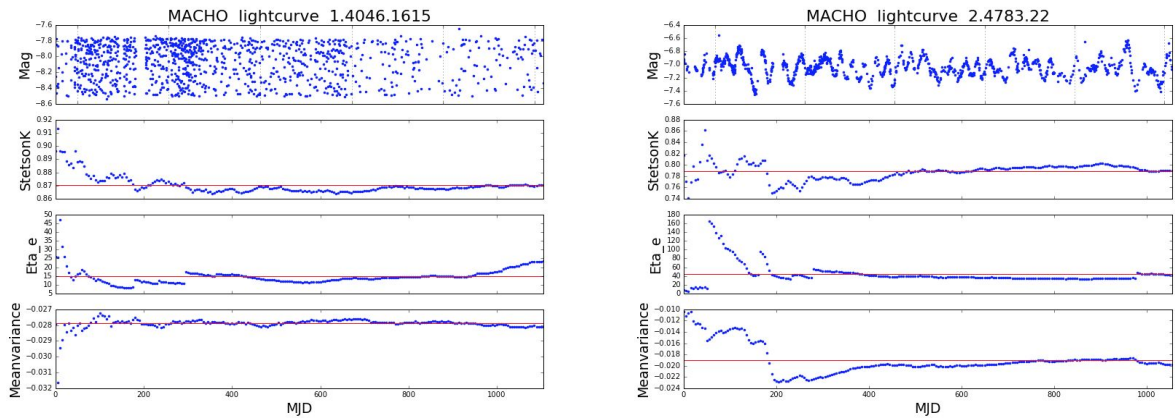

Fig. 1.— Normalized feature values over time. Not all features tend to a specific value (as shown by the mean) as the number of observations increases. Also, not all features converge at the same time.

The implications of this fact are evident: Photometric lightcurves are noisy, non homogeneously measured, with differences in the number of observations among them, and with several observational gaps within them. If the value of the features used to describe them aren't robust and based on long periods of time, then they vary considerably as more observations are added. This kind of features are not reliable to perform classification. In the case of ongoing surveys, the problem is even bigger. The shorter the time series being analyzed, the scarcer the information it contains. In many cases, not even an expert astronomer can correctly classify a lightcurve that consists of only a couple of measurements. This matter is of utmost importance because photometric sky surveys normally take various years to be completed. In cases where the data may not be sufficient to make good use of it, it would be very useful to have a model able to distinguish when there is enough information to make reliable predictions and when there is not. In our work, we focus on generating a model that assigns a level of uncertainty to the calculated features, then the classifier takes it into account and make predictions considering that uncertainty. Features that are calculated over statistical samples are often assigned some measurement of accuracy (Street et al. 1993; Efron & Tibshirani 1994). For simple features like the mean or the standard deviation, closed form equations exist for the associated error. Unfortunately this is not the case for the majority of the time series features used for classification. In the cases where closed form equations are not available, bootstrapping techniques are an adequate alternative (Efron & Tibshirani 1994). This techniques allow to assign measures of accuracy to any statistical quantity by doing random subsampling of the data where it is estimated.

Our model relies on a parametric time series bootstrapping technique, proposed to generate many different lightcurve samples from training sets. Then, various random training sets are built from this samples, where one automatic classifier is trained on each of them. This approach allows to overcome the different biases each training object may possess in its feature values, by averaging over the predictions among different random models.

The objective of this work is to demonstrate the advantages of taking into account the error present in the statistical features used for classification, and show how that error relates to the quality of the time series used for classification. The framework presented in this paper proves that valuable predictions can be made using a small fraction of the observations with which the lightcurves' where originally designed.

The rest of the paper is organized as follows: Section 2 shows a small review of the work done in supervised classification in astronomy and bootstrapping techniques for order dependent data; Section 3 explains the relevant background theory; Section 4 describes the method presented. Section 5 shows the results obtained with the method both for an artificial case and then applied to lightcurve catalogs. Finally, the conclusions are presented in Section 6.

## 2. RELATED WORK

Automatic classification of lightcurves success depends on two important and separated aspects of the process. First is the type of classifier being used. There are many different supervised classification algorithms in machine learning theory, each with its own advantages and limitations. Random Forests (Breiman 2001), Support Vector machines (Cortes & Vapnik 1995), Logistic Regression (Cox 1958) and Decision Trees (Quinlan 1986) are some of the most popular. But no matter which classifier is used, none of them will be successful if the features used for representation are not informative enough and therefore able to distinguish different kind of objects. This is one of the reasons why a lot of the research, regarding automatic classification of variable stars, has focused on the way lightcurves are represented rather than the classifiers they are fed to.

The second aspect is precisely that, how the objects are represented. Lightcurves, being composed of several hundreds of observations, which are hardly ever the same size, unevenly sampled and at different times, are not suited to be introduced to a classifier as input. To address this inconvenient, lightcurves are converted to vectors of numerical values. Great investigations efforts have been made to address this topic, Richards et al. (2011) introduced features that measure different statistical characteristics of time series like: standard deviation, skewness, kurtosis, slopes, and period. Kim et al. (2011a) used features that capture the period, color, amplitude and the autocorrelation value of light curves in order to accurately identify quasars from the MACHO Large Magellanic Cloud database. Also, Pichara et al. (2012) proposed new features based on the parameters of an adjusted continuous autoregressive model of the lightcurves, which generated an improvement in the accuracy of quasar detection methods. Nun et al. (2015) developed a software library which aims to facilitate the feature extraction process. The library includes a very complete compendium of the most important features in recent literature. And because it is open sourced is possible for the whole academic community to ensure that their implementation is correct and to contribute if new descriptors are designed in the future.

Although the techniques previously mentioned are effective, they all rely on features designed by experts in order to describe objects. It is important to mention some efforts to reduce the feature engineering process and contribute to the further automation of the classification process. **Eyer & Blake (2005) propose the use of parameters obtained from a Fourier decomposition of lightcurves in order to classify objects from the All-Sky Automated Survey. In Brett et al. (2004) self-organizing maps are applied to classify singly periodical lightcurves in an unsupervised approach.** In other words their model works directly over unclassified objects, without the need to train on a labeled training set. Finally Mackenzie et al. (2016) proposes an unsupervised feature learning technique, which later can be used to encode and classify variable stars in a supervised fashion.

Regarding Bootstraping methods, they belong to a family of techniques in statistics, that rely on sampling with replacement in order to perform inference (Efron & Tibshirani 1994). They were first introduced by **Efron (1979),** and have become increasingly popular since, because they allow to obtain measures of accuracy (such as the standard error) of a sampling statistic for small samples of data. Their only limitation is that they are computer intensive, because they require to repeat the calculation of the statistics of interest over many bootstrap samples, but the advance that computer power has shown recently makes this easy to overcome and implement.

Because of the previous reasons, and the fact that they can assign deviations measures to almost any statistics, they are perfectly suitable to obtain confidence intervals for the values of complex astronomical features. Nevertheless, the case of lightcurve features is more complicated than usual. Because this descriptors work on time series, which are order dependent data, the manner in which to resample the data is not evident.

Bootstrapping time series, or order dependent data, is not an obvious task, and many different approaches have been proposed along the years. Special considerations must be made, because the data cannot necessarily be changed of order without changing the values of the estimators one wants to calculate. The Block Bootstrap (Kunsch 1989), attempts to solve this issue by dividing the time series observation in adjacent blocks of length $\ell$. Then the resampling is made by drawing this blocks uniformly, thus preserving the original time series structure within each block. Although the choice of $\ell$ is not obvious, Block Bootstrap has been shown to work for general stationary data generating processes (Bühlmann 2002). Kreiss & Franke (1992) introduces a different kind of approach based on autoregressive models and sieve approximation (Grenander 1981). Finally, Kreiss et al. (1998) proposes the so called local bootstrap, which aims to model the dependency that each of observation has on the previous ones. This models proves to be effective only when the observations are generated by a short-range dependent process (Paparoditis & Politis 2000).

Although all of this methods prove to be effective in specific cases, they all make different assumptions over the time series where they are going to be applied, in order to deliver good results (Bühlmann 2002). This, together with the fact that photometric lightcurves do not obey many consistency requirements, make necessary to look for more flexible ways of obtaining bootstrap samples.

## 3. BACKGROUND THEORY

As shown in Figure 1, the value of time series features used for classification fluctuates importantly when the number of observations is small, and normally it tends to converge into more stable values as the lightcurves grow in length. This stabilization process varies for each object and for each of its features. In cases where features need very well sampled time series to be stable, it is harder for a classifier to make accurate predictions. This motivates us to find a method able to assign measures of confidence to the estimated features, and a way for classifiers to adjust their predictions accordingly.

For some simple statistical estimates (like the sample mean for example), closed form equations for the error of the estimate are available. This is not the case for the vast majority of features used in time series classification. The descriptors used in this context, are normally very complex and exact theoretical values can not be obtained. In this kind of cases bootstrapping techniques are an adequate alternative.

In the case of lightcurves, further complications arise. Normal bootstrapping approaches assume that realizations of the random variable are independent of each other, which is not the case of time series data. Lightcurves are measurements of the brightness intensity of an object with time. Therefore, each point is clearly related to the ones nearly observed. In fact, the closer they are the more information they give about each other. Only time-serie-specific bootstrap methods are suitable for this task. Also, the fact that lightcurves are non uniformly sampled, not aligned, have uneven lengths and noisy observations puts even more restrictions to the techniques that might be developed. In this work we pro-pose a Gaussian Process based approach. Gaussian Process (GP) is a very strong and flexible non parametric model that can be used for regression analysis. Because it is non parametric, it works based on a kernel function that defines how any to given observations are related. Several kernel functions can be chosen, depending on how suitable they are for a given problem. In the next section we give further details about Gaussian Processes and its application to time series bootstrapping and bagging in machine learning.

### 3.1. *Gaussian Process regression*

The regression problem corresponds to finding a function $f(\mathbf{x})$ that describes the relation between a vector of input variables $\mathbf{x}$ and a target variable $y$. In practice, however, the process by which data is obtained introduces noise to the values of $y$. In the following review a zero mean **Gaussian** noise on $y$ will be assumed. Therefore:

$$y = f(\mathbf{x}) + \varepsilon, \qquad \varepsilon \sim \mathcal{N}(0, \sigma_n^2)$$

It is important to mention that modern astronomical instruments are normally able to estimate the measurement error $\varepsilon$ associated to each observation. Although this is rarely the case in real applications, it does not affect the concepts presented.

One manner to try and solve the regression problem, and probably the most common one, is to restrict the class of functions for $f(\mathbf{x})$. Then the parameters that govern the model are optimized, so that it fits the observed data as best as possible. This is what is called a parametric approach and, although they are usually easy to interpret, they lack expressive power in more complex scenarios.

Another approach, and the method we use in this paper, is to define a probabilistic model on the functions $f$ that might fit the data, and perform inference directly in the space of functions. This kind of techniques are known as "non parametric Bayesian models" because they establish a prior that reflects the type of functions we expect to see (periodic or soft curves for example), and then make bayesian inference by combining the data that we possess with the prior. This strategy is more flexible because it does not impose any particular type of shape to the curves that might fit the data. Unfortunately, a function, may be evaluated in any number of locations, therefore it is unfeasible to track a probability distribution which describes its values, over a possibly infinitely large input vector $\mathbf{x}$. But, when realizing regression, knowledge over the complete domain of $\mathbf{x}$ is unnecessary. In practice one is only interested in making predictions on a vector $\mathbf{x^*}$ of limited size. This fact make Gaussian Processes able to solve the problem.

Whereas a probability distribution describes the possible outcomes of a random variable (discrete or continuous), a stochastic process governs the properties of functions. A Gaussian Process, in particular, is a collection of random variables, any finite number of which have a joint Gaussian distribution (Rasmussen & Williams 2005). This means **Gaussian Processes** satisfy what is called a marginalization property, which states that if the **Gaussian Process** specifies $(y_1, y_2) \sim \mathcal{N}(\mu, \Sigma)$, then it must also specify $y_1 \sim \mathcal{N}(\mu_1, \Sigma_{11})$. In other words, if it implies a distribution over a (possibly infinite) set of variables, then that same distribution applies for a smaller set of those variables. Therefore this property allows to make the same inference as if one was dealing with the infinite set of variables, when only working with the ones that are of interest.

A **Gaussian Process** is completely defined by its mean and covariance functions $m(x)$ and $k(x, x')$. On one hand, the mean function specifies the general tendency of the functions that will arise. To make an example, in many real applications the mean function is simply defined as $m(x) = 0$. Which means the average value of the functions perceived, at any given point $x$, is 0. On the other hand, the covariance function $k(x, x')$ defines the shape of the curves that appear, by determining the covariance between any two points. More formally, the mean and covariance functions that govern a real process $f(\mathbf{x})$ are:

$$m(\mathbf{x}) = \mathbb{E}[f(\mathbf{x})],$$
$$k(x, x') = \mathbb{E}[(f(\mathbf{x}) - m(\mathbf{x}))(f(\mathbf{x}') - m(\mathbf{x}'))]$$

and the Gaussian Process

$$\mathbf{x} \sim \mathcal{GP}(m(\mathbf{x}), k(\mathbf{x}, \mathbf{x}')).$$

Then in order to sample functions from a **Gaussian Process** prior, one must simply build a multinomial **Gaussian** distribution, by replacing the $x_*$ where one wants to sample in the mean function, and covariance function of choice, and with that build the corresponding $\mu(x_*)$, and $\Sigma(x_*)$. Assuming $m(x) = 0$, and a number of input points $X_*$ then the function evaluated at those points $f_*$ satisfies:

$$\mathbf{f}_* \sim \mathcal{N}(0, K(X_*, X_*)).$$

To further increase the understanding, **let us assume** a Gaussian Process prior with a mean function $\mu(x) = 0$ and the following kernel function:

$$\mathrm{cov}(f(\mathbf{x}_p), f(\mathbf{x}_q)) = k(\mathbf{x}_p, \mathbf{x}_q) = \exp(-\frac{1}{2}|\mathbf{x}_p - \mathbf{x}_q|^2).$$

This function is called squared exponential and is one of the most common kernel functions. Figure 2 shows three samples taken at random from this prior.

Finally, having assumed a given GP prior one must be able to incorporate the information the training data provides from the phenomenon. In bayesian terms this corresponds to combine the likelihood of the functions, given the observed points, with the prior that has been chosen, in order to get the posterior distribution. The joint distribution of the training outputs $\mathbf{f}$ and the test outputs $\mathbf{f}_*$ according to the prior is

$$\begin{bmatrix} \mathbf{f} \\ \mathbf{f}_* \end{bmatrix} \sim \mathcal{N}\left(0, \begin{bmatrix} K(X, X) & K(X, X_*) \\ K(X_*, X) & K(X_*, X_*) \end{bmatrix}\right)$$

To get the posterior distribution, the joint distribution must be conditioned to produce only those functions that

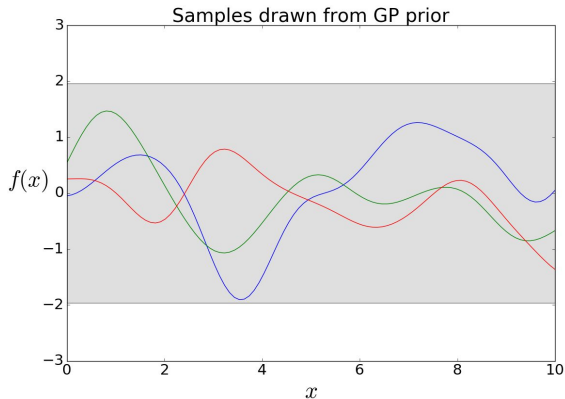

Fig. 2.— Three sampled functions from a GP prior with squared exponential kernel function.

are consistent with the observed data points. This becomes simply

$$\mathbf{f}_*|X_*, X, \mathbf{f} \sim$$
$$\mathcal{N}\big(K(X_*, X)K(X, X)^{-1}\mathbf{f},$$
$$K(X_*, X_*) - K(X_*, X)K(X, X)^{-1}K(X, X_*)\big)$$

Now, in real cases where observations are noisy, this equations can very easily be updated to incorporate this deviations. The covariance function, regardless of the one that is being used must be updated to

$$\operatorname{cov}(y_p, y_q) = k(\mathbf{x}_p, \mathbf{x}_q) + \sigma_n^2 \delta_{pq}$$
$$or$$
$$\operatorname{cov}(\mathbf{y}) = K(X, X) + \sigma_n^2 I$$

where $\delta_{pq}$ is the Kronecker delta. Then the joint distribution becomes:

$$\begin{bmatrix} \mathbf{y} \\ \mathbf{f}_* \end{bmatrix} \sim \mathcal{N}\left( 0, \begin{bmatrix} K(X, X) + \sigma_n^2 I & K(X, X_*) \\ K(X_*, X) & K(X_*, X_*) \end{bmatrix} \right)$$

and one can finally arrive to the key predictive equations for Gaussian Process regression.

$$\mathbf{f}_*|X, \mathbf{y}, X_* \sim \mathcal{N}\left( \bar{\mathbf{f}}_*, \operatorname{cov}(\mathbf{f}_*) \right),$$
$$\text{where}$$
$$\bar{\mathbf{f}}_* \triangleq \mathbb{E}[\mathbf{f}_*|X, \mathbf{y}, X_*]$$
$$= K(X_*, X)[K(X, X) + \sigma_n^2 I]^{-1}\mathbf{y},$$
$$\operatorname{cov}(\mathbf{f}_*) = K(X_*, X_*) - K(X_*, X)[K(X, X)$$
$$+ \sigma_n^2 I]^{-1}K(X, X_*).$$

For the complete derivation of this equations please refer to Rasmussen & Williams (2005).

In the case of regression problems the mean of the distribution formed by the posterior is taken as the function that best represents the relation between the input and the objective variable. One of the main advantages of this regression model, other than its flexibility, is that it does not only gives the values of the function evaluated on some locations $X_*$, but also, because it is probabilistic, the prediction has a deviation assigned to it. As Figure 3 shows, this deviation reflects very accurately the knowledge data provides. As it tends to be smaller near the data points, and grows in the intervals where there are not any observations.

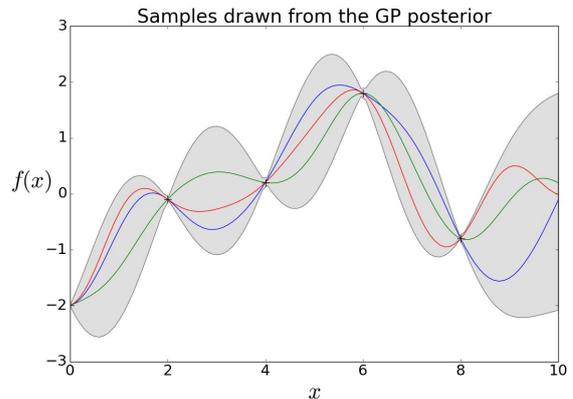

Fig. 3.— Three sampled functions from the GP posterior conditioned on five observations. The standard deviation is smaller close to the observations and gets bigger as one moves away.

### 3.2. Gaussian Process bootstrap

Because the Gaussian Process is a probabilistic model, it can be used in other ways rather than just a simple regressor. Kirk & Stumpf (2009), shows an example of how one can apply Gaussian Process regression to form a parametric time series bootstrap. The technique is straight forward. A GP is adjusted over the time series of interest and the posterior distribution that best explains the behavior of the data is obtained. **As shown in Section 3.1, the posterior is a Multivariate Gaussian Distribution completely defined by its mean vector $\bar{\mathbf{f}}_*$ and covariance matrix $\operatorname{cov}(\bar{\mathbf{f}}_*)$. Then, several possible time series can be randomly sampled from this distribution, until a sample set of the desired size is formed.**

This has many advantages over more traditional bootstrap approaches. First, it takes into account the relation different observations have on each other, and their relative position in the curve. In other words, if an observation is being sampled from an isolated fragment of a series, the value will vary considerably across different samples. While samples that have actual observations near them, will have similar values to the points around them. Second, the way observations influence each other can be controlled depending on the kernel function that is chosen. If periodic relations are expected or seen in the data, a periodic term can be added to the kernel for example. Third, depending on the kernel that is used it allows to take into consideration the error in the values of the data that one possess. Fortunately, in the case of photometric lightcurves, catalogs possess the measurement error for every observation. Therefore this information can be added to the model in order to increase its accuracy, because the model knows beforehand which data points are more reliable than the rest. Finally, it uses all the observations available to create the sampled curves, whereas other bootstrap techniques work by dividing the data into subsets where valuable information may be lost.

### 3.3. *Bagging*

Bagging stands for bootstrap aggregating and is a machine learning ensemble strategy first introduced by Breiman (1996). It allows to combine the strength of multiple models in order to increase the overall predicting accuracy. The idea behind bagging is to generate many versions of the same predictor, where each version is trained on a different bootstrap sample of the original training set. Then, in the case of objects classification, the most voted class among the group of models is regarded as the final output. Bagging not only improves the predictive power of the models, but also, by taking the voting distribution, it gives a confidence measure of the prediction it makes.

Bagging is specially effective when the predictive method presents a high instability. Büchlmann & Yu (2002) formalize the notion of instability and show how this technique helps to overcomes the effects it has in classification performance. The formal mathematical definition escapes the scope of this paper, but the general idea is that instability is bigger when the model being adjusted does not converge to a definite value after a certain amount of data. In other words, if small changes in the data considered to train, or new observations of the same, produce differences in the final model. This is precisely the case shown in Figure 1. If the value of the features is highly unstable due to the small amount of observations, then the learned model will suffer the same problem, and the predictions it realizes will not be reliable.

### 4. METHODOLOGY

As demonstrated before in Section 1, when lightcurves are composed of only a few points the value of the features that describe them becomes disperse. This because as there are little observations the value of each one becomes more important, and tiny variations on their values, or the presence of new ones, affects the estimation considerably. This deteriorates the effectiveness of classifiers as features are not longer able to describe different objects consistently. To overcome this problem we draw from what is proposed in Kirk & Stumpf (2009), to create bootstrapped samples of any feature, together with a bagging approach to combine the different outcomes each set of samples produces. By doing this we diminish the effects that feature variance has on classification performance.

Our algorithm consists of four major steps. In the first stage, we adjust a Gaussian Process regression model to each lightcurve. **This produces a probability distribution, for each individual curve, that represents the different values that curve may take. Then $n$ time series are randomly sampled from each of this models, according to the technique described in Section 3.2.** In the second stage we take a different sample from each of the original objects to form $n$ different sets. Then we calculate a set of descriptors for each of the samples in this so called "sample sets". The third stage consists of training a classifier on each of this sets, thus obtaining $n$ different models. The fourth and final step is to classify the unknown lightcurves. For this we use the same idea, $n$ samples are taken from the adjusted GP on the lightcurve. Finally each of this samples is classified by one of the models, thus obtaining a voting distribution on each object's class. Figures 4 to 8 show the different stages of the process.

### 4.1. *Time series bootstrapping*

The first step of the process is to take every lightcurve in the training set, and take bootstrapped samples from each of them. The idea is that each of this lightcurve presents different behaviors in the sections where the sampling is poor, in other words where not many measurements where made. On the contrary if the lightcurve presents a very good sampling we expect the bootstrapped samples to be very similar.

To obtain bootstrap samples of the lightcurves in the training set we adjust a Gaussian Process regression model on each of them, and take $n$ samples from the obtained posterior distribution. As described in Section 3.1, what defines the shape of a Gaussian Process is the kernel function. In the case of photometric lightcurves, we take from the work done by Faraway et al. (2014), and use a similar Gaussian Process prior to the one they proposed. Although, because the type of object is not known beforehand, we assume a constant mean function equal to the average value of each signal. Then the prior we use is:

$$f(\mathbf{x}) \sim \mathcal{GP}\left(\mu(\mathbf{x}), k(\mathbf{x}, \mathbf{x}')\right)$$

where

$$\mu(\mathbf{x}) = \frac{1}{n}\sum_{i=1}^{n}\mathbf{x}_i$$

and

$$k(\mathbf{x}_p, \mathbf{x}_q) = \sigma_f^2 \exp\left(-\frac{1}{2l^2}(\mathbf{x}_p - \mathbf{x}_q)^2\right) + \sigma_n^2 \delta_{pq}$$

In the equations above, $\mu(\mathbf{x})$ is the mean of the signal, $\sigma_f^2$ is the signal variance, $l$ is the length scale, $\delta_{pq}$ is the Kronecker delta and $\sigma_n^2$ is the noise variance. The last term is particularly interesting, because for astronomical data, unlike the majority of cases, the measurement error can be estimated for each observation. Figure 5, shows the adjusted Gaussian Process model over a lightcurve from the MACHO catalog.

The number of samples to take is not obvious at first hand and, because it may change in different scenarios, it must be found empirically. There is a trade-off between the accurate representation of the curves distribution and the computational time the method takes. In our experiments we found that 100 samples gave optimal results while still being computationally feasible. Figure 6 shows two random samples taken from a GP model adjusted over the curve on Figure 5.

**Time series Bootstrap**

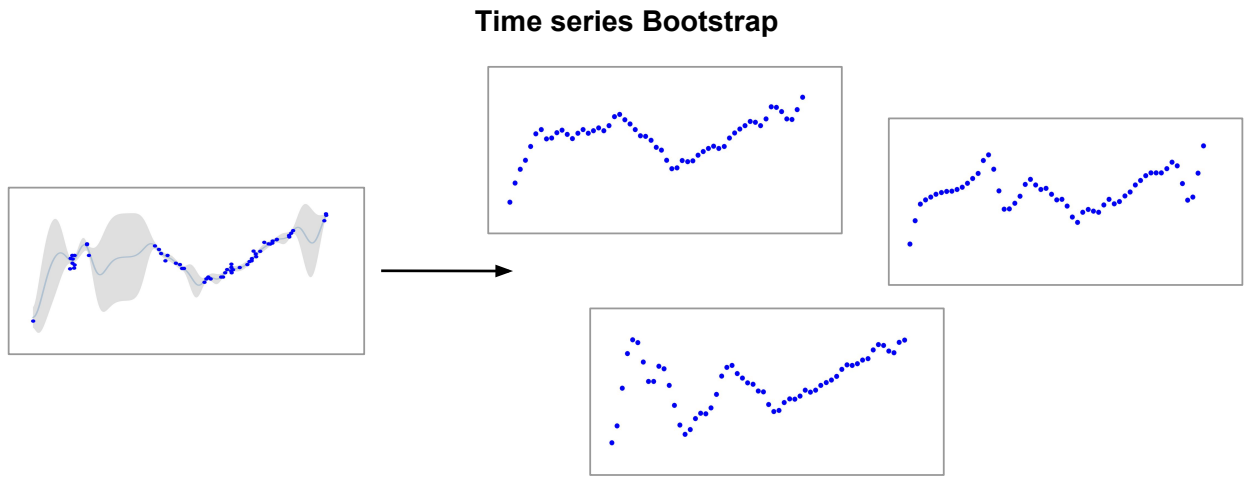

FIG. 4.— Illustration of the first stage of the algorithm, the Time series Bootstrapping. A Gaussian Process is adjusted over a lightcurve and several random sampled curves are obtained it.

### 4.2. *Sample sets*

After taking the bootstrap samples, we form $n$ different training sets. Where each set contains a single and different sample for each of the original labeled lightcurves. We refer to this sets as the sampled sets. An illustration of this stage is shown in Figure 7.

Then a group of time series features is calculated for each curve of the sampled sets. For this task we use FATS (Nun et al. 2015). This open sourced python library allows easy and efficient calculation of the most used lightcurve features existent in literature. Although this tool allows to calculate more than 50 different time series features, we restricted our work to a subset of only twenty three features that prove to be effective for classification. We decided to discard all features that need different bands to be calculated, because this adds further complexity to the problem, and including them goes beyond the scope of this investigation.

At this stage, because the features have been calculated for $n$ bootstrapped samples of each lightcurve, we now possess an estimation of the distribution of their values for each object. According to this distributions, features that present a high variability in their values will be less influential on the classification, whereas features that are more consistent will be taken more into account by the model.

### 4.3. *Training*

After we calculate the features for each of the samples sets, we adjust one decision tree classifier (Breiman et al. 1984; Quinlan 1986) on each sample set. We decide to use decision trees on the samples in order to form a Random Forest (RF) Classifier (Breiman 2001) when we ensemble the trees. RF has proven to be one of the most effective classifiers for variable star classification (Carliles et al. 2010; Richards et al. 2011; Pichara et al. 2012; Pichara & Protopapas 2013). Although instead of combining trees, trained with different subsets of features, we combine trees trained on different random scenarios. Each scenario is a possible uncertain outcome of the values of the original training set.

### 4.4. *Classification*

The final stage is to predict the class of a new unlabeled object. For this the same logic presented before is used. Because the values of a new lightcurve may be corrupted, the prediction yielded by the classifiers have a greater chance of being incorrect. Therefore, again, $n$ different samples are obtained and their features calculated. Then each of this samples is given to a different trained model for it to cast its vote. Finally the vote of all models is combined and the most popular class is regarded as the final predicted class. An illustration of this stage is shown in Figure 8.

It is important to note, that because a voting is taken place, the actual prediction of this framework gives a belief of belonging to each of the possible classes. One can take advantage of this quality to discard, or further analyze confusing results, in the case, for example, that many models give different predictions regarding the same object.

If a lightcurve presents very little, noisy, or unevenly distributed measurements, the value of its features will change greatly among different samples. Therefore it is likely for different classifiers to be confused and cast contradicting votes. On the contrary, if a lightcurve is well sampled, and thus very well described, the voting of the different classifiers is likely to be more consistent.

### 5. EXPERIMENTAL RESULTS

In this section the experimental results are presented. First, we detail a synthetic experiment based on the Robot navigation dataset. The goal of this example is to show how classification results are affected when the value of the variables are affected by randomness. And then, how this problem can be reduced by using bagging techniques like the one proposed. Then we present the classification results obtained by working with photometric lightcurve data. In order to generate an experimental setup for the problem of automatic classification with incomplete lightcurves, the lightcurves are truncated into smaller versions of themselves by selecting only the first few observations. This way we simulate the scenario of surveys that are barely beginning their measurement process. The difference between the real

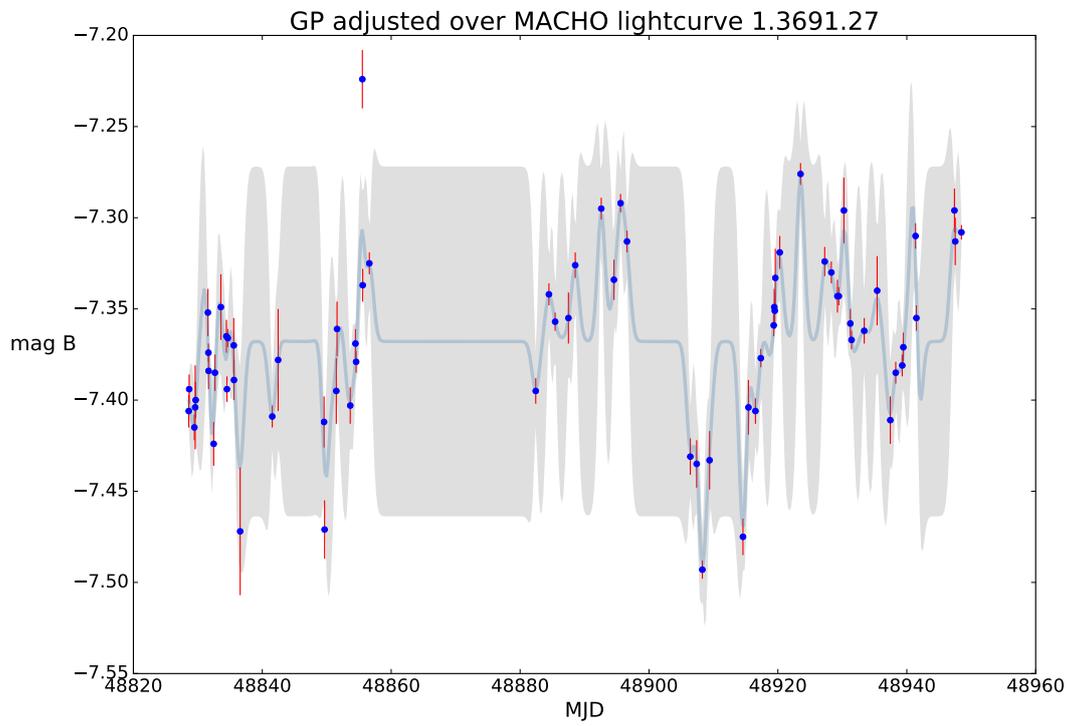

Fɪɢ. 5.— Gaussian Process regressor adjusted over a lightcurve from the MACHO catalog. The model captures the general form of the time series, and adjusts the deviation according to the observations possessed. The model is less influenced by measurements with greater measurement error.

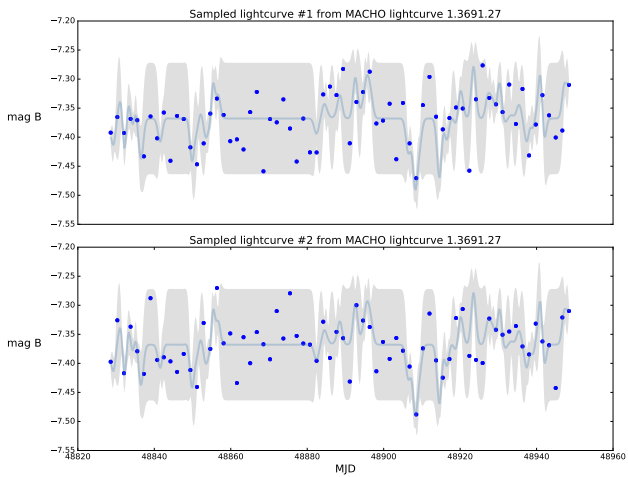

Fig. 6.— Two random samples taken from a GP model of a MA-CHO lightcurve. The samples are taken at uniform times over the span of the measurements. Sampled observations near the original ones have very similar values, while samples taken from empty spaces are more disperse.

| | Class | Number of Objects |
|---|---|---|
| 1 | Move-Forward | 2205 |
| 2 | Sharp-Right-Turn | 2097 |
| 3 | Slight-Right-Turn | 826 |
| 4 | Slight-Left-Turn | 328 |

TABLE 1: Robot Training Set Composition

case and the synthetic one (and one of the key contributions of this investigation) is how the method proposed in section 3.2 is used to obtain the bootstrapped samples of noisy lightcurve features. In both the synthetic and real cases we compare how a bagging scheme classifier improves the classification of standard models. Classifier performance is measured with a 10-fold stratified cross-validation F-Score on each of the classes present in the corresponding data set. We choose the classic Decision Tree (Breiman et al. 1984; Quinlan 1986, 1996) and the Random Forest (Breiman 2001) as the classifiers which to compare our model with. We compare with the Decision Tree to validate that the bagging realized in our method improves the results of this simple model. Second, we compare with the Random Forest because this is the classifier of choice in many recent literature (Nun et al. 2014; Kim et al. 2011a; Pichara et al. 2012; Pichara & Protopapas 2013; Richards et al. 2011) regarding automatic classification of variable stars, and is also the most precise according to our tests. All three models work **with** the exact same set of features.

### 5.1. Robot dataset

The dataset used for this experiment is taken from the UCI machine learning repository (Lichman 2013). It is called "wall following robot" dataset (Freire et al. 2009) as it was collected from a mobile robot which navigates along the walls of a room without colliding. The robot is equipped with a belt of 24 ultrasound sensors that measure the proximity of objects in a 360 degree radius at evenly timed steps. Then each entry of the dataset contains the readings of the 24 sensors together with a class, which corresponds to the specific movement the robot must make, from a group of four defined possible movements.

The robot's training set is composed of 5456 readings, and the class composition is detailed in Table 1. We choose to work on this dataset as a synthetic example, because it does not have any missing values and it also has a similar number of attributes and instances as the photometric datasets we work with.

To evaluate the effects that feature noise has in classification results **we first test the performance of two regular classifiers over a normal dataset with increasing levels of noise in its variables. Then, this results are compared with an ensemble of classifiers working on synthetic bootstrapped samples of the same training set. The experiment is the following.**

The robot dataset is taken and the range of each feature is calculated (the difference between the maximum and the minimum value it takes on the dataset). Then to each feature, of each instance, a white noise kernel is added, with standard deviation equal to a randomly chosen value between zero and a fixed percentage of the amplitude. Hence, for example, to generate a dataset with a 5% of noise, we take a sample from all of those kernels by using 5% of the corresponding feature range as the maximum possible standard deviation. The advantage of doing this is that it allows us to generate any number of randomly sampled sets from the same feature distribution.

**It is important to choose a different white noise kernel, for a same feature, on different instances. This way the obtained datasets will resemble the ones used for photometric classification. Where, as shown in Figure 1 a same variable may behave differently on different lightcurves.**

**The regular classifiers (a Decision Tree and a Random Forest) are trained on a single dataset, for each level of noise. Whereas our method, as described in Sections 4.3 and 4.4 trains a model on each different sampled set and then uses a voting scheme classification in order to make predictions.**

We did this test for various levels of added noise, ranging from 5% to 20%. The results obtained are shown in Figure 9. We can see how for all of the classes, the voting scheme classifier gives better results than both the Decision Tree and the Random Forest, trained over a single observed random set.

### 5.2. MACHO dataset

The MACHO catalog is the result of a project that aimed to find dark matter in the form of massive compact halo objects (MACHOs). The project made photometric observations of tens of millions of stars, for almost 6 years, in the Large Magellanic Cloud (LMC), Small Magellanic Cloud (SMC) and Galactic bulge (Alcock et al. 2001).

The photometric training sets are labeled subsets of the actual surveys. The MACHO training set is composed of 6627 curves (Kim et al. 2011b). Its class composition is detailed in table 2.

**Sample sets**

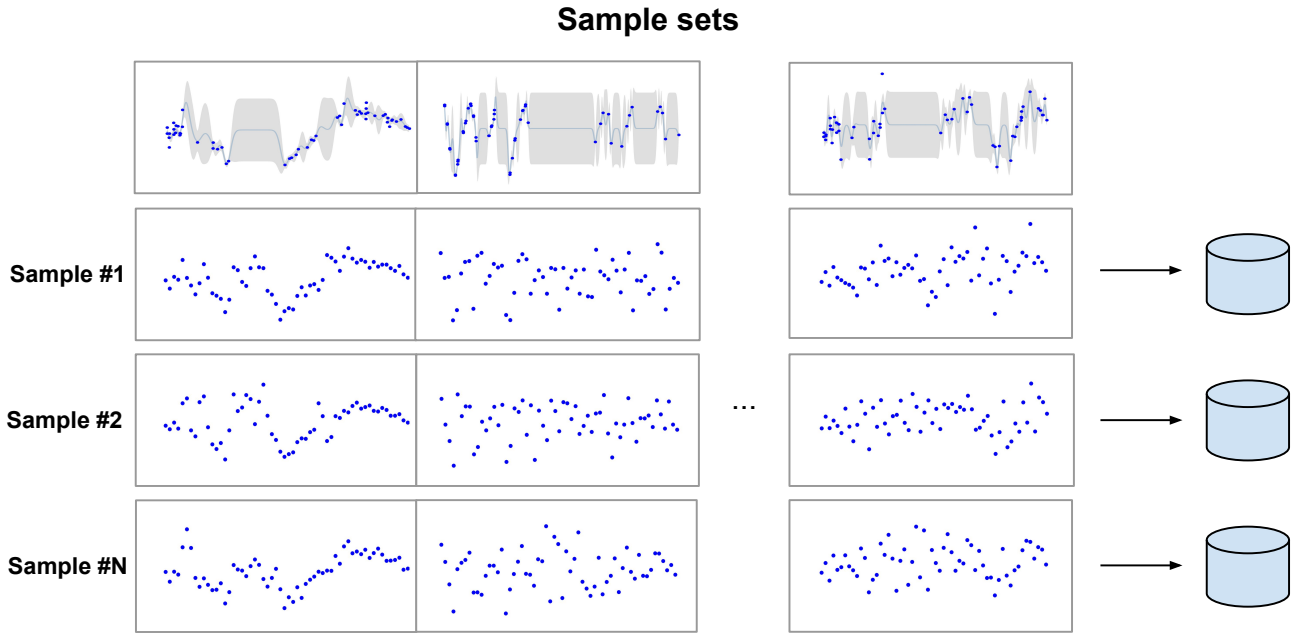

FIG. 7.— Illustration of the second stage of the algorithm. The different samples of each lightcurve are separated into different "sample sets". Each of this sets represents a different random scenario of the observed lightcurves.

**Classification**

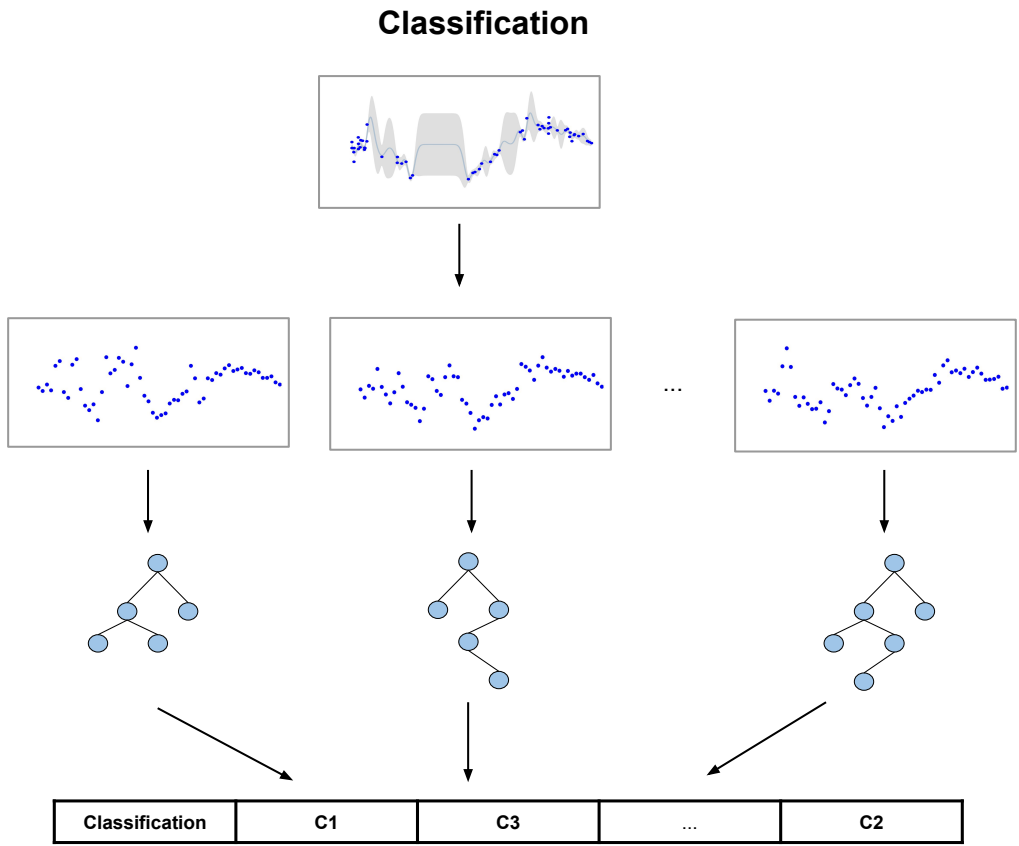

FIG. 8.— Illustration of the final stage of the algorithm. When an unknown lightcurve needs to be classified, the same process is realized. Various samples are taken from it, their features calculated, and then given to a different classifier from the ones trained on the previous step.

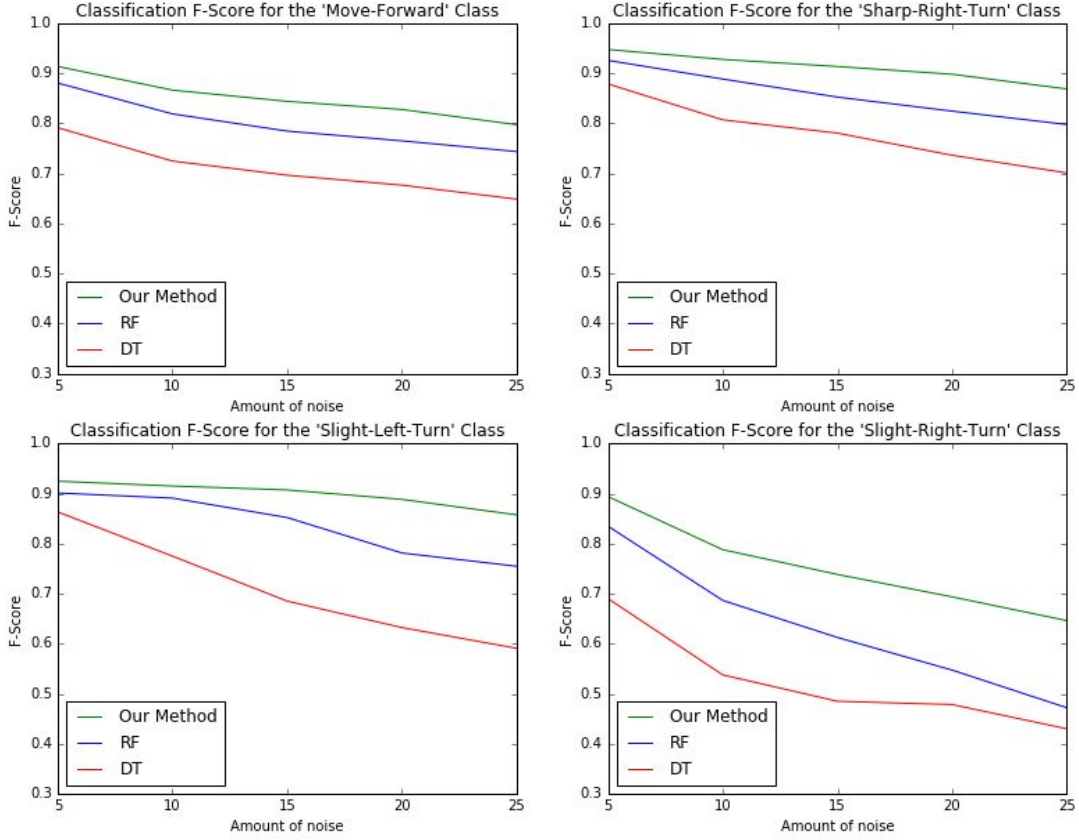

Fig. 9.— Classification F-Score for the Robot training set. The results obtained by bagging the predictions of many different classifiers are less affected by noise than both the Decision Tree and the Random Forest.

|   | Class | Number of Objects |
|---|---|---|
| 1 | Non Variable | 4768 |
| 2 | Quasar | 34 |
| 3 | Be Star | 112 |
| 4 | Cepheid | 101 |
| 5 | RR Lyrae | 606 |
| 6 | Eclipsing Binary | 255 |
| 7 | MicroLensing | 393 |
| 8 | Long Period Variable | 358 |

TABLE 2: MACHO Training Set Composition

Figure 10 shows a Gaussian Process model adjusted over a lightcurve from the MACHO catalog. It is important to notice that the model assigns greater uncertainty to regions where no observations are recorded, while regions with better measurements are regarded as more accurate. This is very important, because lightcurves with greater gaps in their measurements will produce bootstrapped samples with greater differences in their values, while better sampled curves will result in more consistent ones.

Figure 11 shows three samples taken randomly from the previous model. It is evident that all the samples present very similar values on regions with higher density of observations. On the other hand, regions where the original time series has less information, are very different among the samples. This is the behavior expected for this stage of the process.

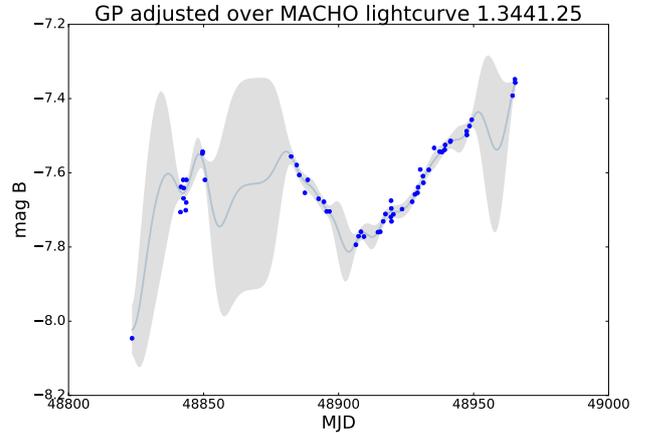

Fig. 10.— Gaussian Process regressor adjusted over a lightcurve from the MACHO catalog. The model gives greater uncertainty to regions where no observations are recorded.

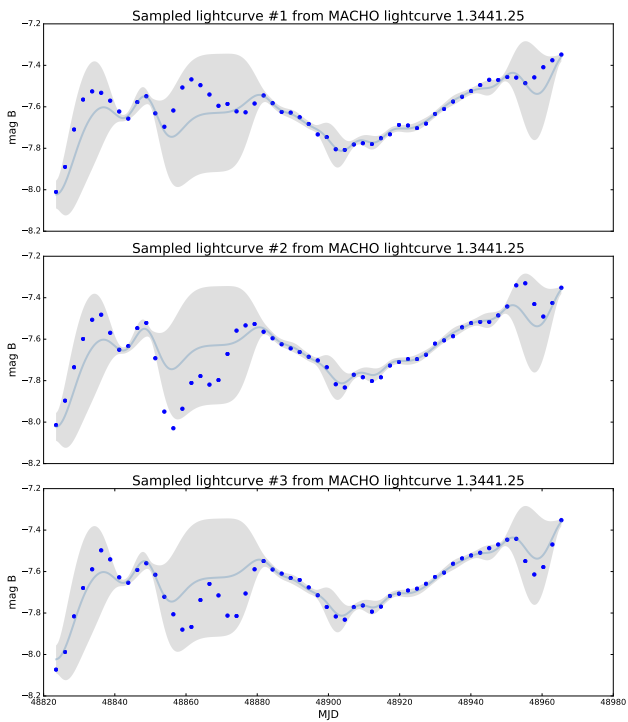

Fig. 11.— Three random samples taken from a GP model of a MACHO lightcurve. Samples taken from empty spaces are more disperse. Therefore lightcurve with poorer sampling, both in total number and uniformity of observations, will present a greater dispersion in the value of their calculated features.

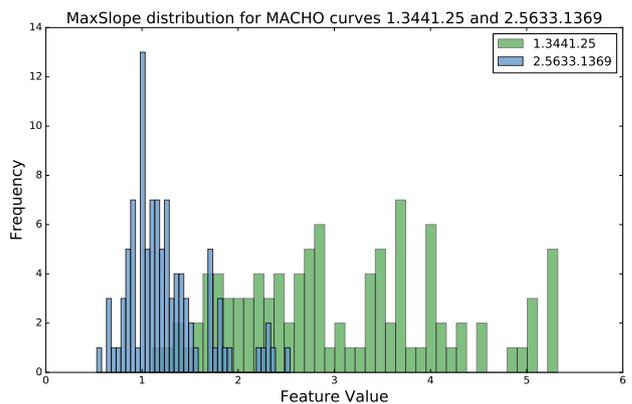

Fig. 12.— Distribution for the values of the MaxSlope variability measure for two lightcurves from the MACHO catalog. It is evident that the blue values are more concentrated and thus present lower variability. On the other hand the green values show a bigger error in the estimation of its value.

Every statistical estimate has an inevitable degree of error in its estimation. Therefore finding methods to assign measures of accuracy in their values is crucial. Variables which values present high degrees of error (just as some photometrical measurements) are normally dismissed versus more precise ones when using them for analyses. The bootstrapping technique used in this investigation allows for the same logic to be applied to the time series features used for classification. Figure 12 shows a graphical comparison of the distribution of the MaxSlope variability feature for two different curves from the MACHO catalog. The MaxSlope corresponds to the maximum absolute magnitude slope, between two consecutive observations, present on the represented series of points. It is evident that one curve presents much more error in the estimation of the MaxSlope variability feature.

The curve that presents more consistency in its values, will be more influential in the classification process than the other one. Because as the values will be given to different classifiers, inconsistent behaviors are dismissed by the voting of the majority, while consistent ones are reinforced.

Table 3 show the classification results obtained by the model proposed in this paper, a Random Forest and a Decision Tree, applied to the MACHO training set. Compared with the Decision Tree, our method shows better results for all of the classes training set, except for the Cepheids. These results show that combining the votes of many decision trees, over different samples of the same objects, effectively improves the classification performance.

Compared with the Random Forest, although there are specific differences on the per class performance, both models have similar results on the MACHO training set. Our method gets better results for RR Lyrae stars and Quasars, while the RF does better at identifying Cepheids and Microlensings.

| | Class | Random Forest | Our Method | Decision Tree |
|---|---|---|---|---|
| 1 | Be Star | 0.570 | 0.546 | 0.461 |
| 2 | Cepheid | 0.931 | 0.790 | 0.870 |
| 3 | Eclipsing Binary | 0.474 | 0.465 | 0.392 |
| 4 | Long Period Variable | 0.877 | 0.856 | 0.850 |
| 5 | RR Lyrae | 0.737 | 0.762 | 0.671 |
| 6 | Microlensing | 0.823 | 0.775 | 0.690 |
| 6 | Non Variable | 0.930 | 0.936 | 0.910 |
| 6 | Quasar | 0.041 | 0.247 | 0.130 |

TABLE 3: Classification F-Score on the MACHO training set

### 5.3. OGLE-III dataset

The OGLE-III catalog of variable stars (Udalski et al. 2008) contains photometric data obtained during the third phase of The Optical Gravitational Lensing Experiment. This wide-field sky survey was designed with the objective of finding dark matter through the microlensing technique. It contains regular measurements of the brightness of more than 200 million objects, from the large and small Magellanic Clouds, the Galactic bulge and the Galactic Disk, taken since 2001.

The OGLE training set is composed of 4733 labeled curves. Its class composition is detailed in table 4. This training set is chosen as a subset of the most represented variable star classes in the catalog, and of comparable size to the MACHO dataset. Also, in order to make the classification more difficult, we choose objects from different locations in space, namely the Large Magellanic Cloud, Small Magellanic Cloud, and the Galactic Disk.

| | Class | Number of Objects |
|---|---|---|
| 1 | Cepheid | 724 |
| 2 | Type 2 Cepheid | 575 |
| 3 | RR Lyrae | 998 |
| 4 | Eclipsing Binary | 794 |
| 5 | Delta Scuti | 656 |
| 6 | Long Period Variable | 986 |

TABLE 4: OGLE-III Training Set Composition

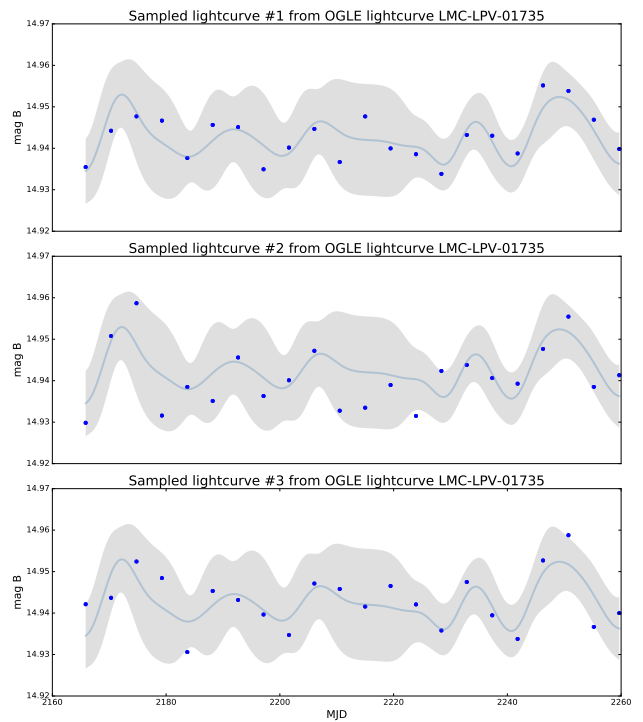

Fig. 14.— Three random samples taken from a GP model of an OGLE lightcurve. Due to the higher instrumental error this observations present, the sampled lightcurves present considerable dispersion everywhere.

Figure 13 shows the fitting of the Gaussian Process model over a lightcurve from the OGLE catalog. The model again is able to describe the general behavior of the curve but this time it shows a greater dispersion along most of the curve. This result, as described in section 4.1, is the effect of the noise variance component of the kernel used by the Gaussian Process. Because the observations of this curve present a higher measurement error, the model automatically assume the regression must not fit that close to those observations.

Figure 14 shows three random samples taken from the fitted model. In this case, due to the general dispersion of the model, the samples tend to be more different from one another.

Figure 15 this time shows the distribution of the $\eta^e$ variability feature (Kim et al. 2014) but for two lightcurves from the OGLE catalog. Again we can see that the difference in variability of the values for the two curves is considerable.

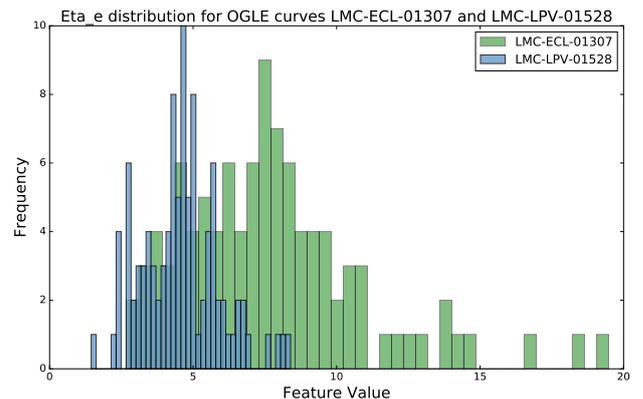

Fig. 15.— Distribution for the values of the eta variability measure for two lightcurves from the OGLE catalog. It is evident that the the blue values are much more concentrated and thus present lower variability. On the other hand the green values show many escaped higher values.

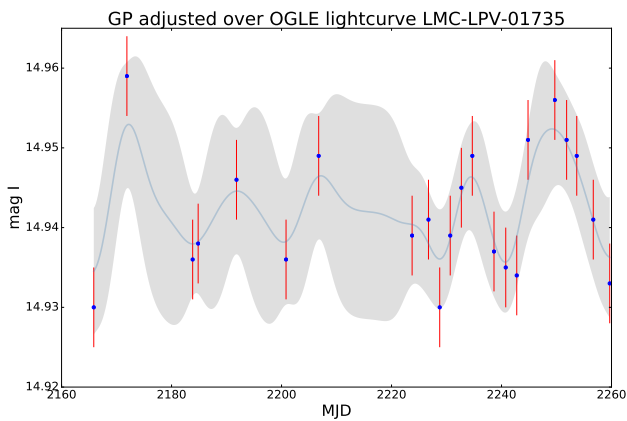

Fig. 13.— Gaussian Process regressor adjusted over a lightcurve from the OGLE catalog. When observations contain higher instrumental error, the model assigns more dispersion to the general distribution.

Table 5 show the classification results obtained by the model, but this time working on the OGLE training set. The results obtained are very similar to the ones shown in MACHO dataset. When compared with the Decision Tree, except for the Eclipsing Binaries, all classes see their F-Score improved by our model. This again vali-dates the potential of the model.

Compared with the Random Forest the results again are similar with the difference that our method gives better results for Cepheids and RR Lyrae stars, which are extremely valuable to find, compared with the rest of the classes.

Although the proposed model does not outperforms the random forest classifier is important to notice the high precision the model presents for some important variability classes. For example, RR Lyrae stars have an 0.89 F-Score which is really good, even though the model is only working with five percent of the available observations. Long period variables are even more impressive with a 0.96 F-Score. These results show that astronomers may not need to wait long periods of time to identify this type of objects reliably.

|   | Class | Random Forest | Our Method | Decision Tree |
|---|---|---|---|---|
| 1 | Cepheid | 0.804 | 0.833 | 0.757 |
| 2 | Delta scuti | 0.824 | 0.825 | 0.807 |
| 3 | Eclipsing Binary | 0.872 | 0.728 | 0.845 |
| 4 | Long Period Variable | 0.974 | 0.963 | 0.954 |
| 5 | RR Lyrae | 0.832 | 0.891 | 0.704 |
| 6 | Type II Ceph | 0.775 | 0.785 | 0.694 |

TABLE 5: Classification F-Score on the OGLE training set

## 6. CONCLUSIONS

In this work, we present a new way of bootstrapping features for lightcurve classification where, instead of making subsamples of the instances of the training set, we sample the original time series used to estimate them.

A Gaussian Process Regression is used to form a probabilistic model of the values observed for each lightcurve. In bayesian terms, this is called a posterior distribution, because it combines the evidence the data gives, with a prior that reflects the beliefs we have on the behavior of stellar variability. The prior also considers the measurement error each observation presents and adjusts the model accordingly. We performed tests on the MACHO and OGLE catalogs and our results show that the regression model correctly describes the behavior of the lightcurves. Because the Gaussian Process is a generative model, it uses all of the observations to form new samples, instead of only considering the information of preceding points. This preserves the long term patterns underlying in the data. The model also assigns greater deviation to the regions where no observations are recorded. Therefore samples taken from empty spaces are more disperse than the ones taken near other observed points. We have also shown how to obtain an empirical distribution of the value of any feature. Lightcurves with poorer sampling, both in total number or uniformity of observations, present a greater dispersion in the value of their calculated features. This allow for a model to discard the instances which values have higher variability for others with more consistency in their values. We show that combining the votes of many different classifiers across different samples of the same objects increases the overall classification accuracy. Although it does not outperforms the random forest classifier on every class, both models show that they are able to recognize some classes with surprising precision, in spite of working with only a fraction of the observations. Finally we have shown that our method is able to classify some variability classes with only a fraction of the observations of the original lightcurve. For example, RR Lyrae Stars and **Long Period Variables** can be identified with more than 80% of accuracy using only the first 5% of the observations available in MACHO and OGLE. This could allow better utilization of the early stages of survey exploration. We believe this framework constitutes the first attempt to include the error of time series features into the automatic classification process. In this sense, it proves that better results can be obtained by using simpler models, like decision trees, when this issue is taken into account. We hope that this research encourages the astronomical community to take more into consideration the error associated with feature calculation, how the measurement process impacts it, and how to develop more ways to overcome it.

### ACKNOWLEDGMENTS

This work is supported by Vicerrectoría de Investigación (VRI) from Pontificia Universidad Católica de Chile, the Institute of Applied Computer Science at Harvard University, and CONICYT-Chile, through the FONDECYT project number 11140643.

REFERENCES

Alcock, C., Allsman, R., Alves, D. R., Axelrod, T., Becker, A. C., Bennett, D., Cook, K. H., Drake, A. J., Freeman, K., Geha, M., et al. 2001, The Astrophysical Journal Supplement Series, 136, 439

Bloom, J., & Richards, J. 2011, Advances in Machine Learning and Data Mining for Astronomy, 89

Bloom, J., Richards, J., Nugent, P., Quimby, R., Kasliwal, M., Starr, D., Poznanski, D., Ofek, E., Cenko, S., Butler, N., et al. 2012, Publications of the Astronomical Society of the Pacific, 124, 1175

Breiman, L. 1996, Machine learning, 24, 123

—. 2001, Machine learning, 45, 5

Breiman, L., Friedman, J., Stone, C. J., & Olshen, R. A. 1984, Classification and regression trees (CRC press)

Brett, D. R., West, R. G., & Wheatley, P. J. 2004, Monthly Notices of the Royal Astronomical Society, 353, 369

Bühlmann, P., & Yu, B. 2002, Annals of Statistics, 30, 927

Bühlmann, P. 2002, Statistical Science, 52

Carliles, S., Budavári, T., Heinis, S., Priebe, C., & Szalay, A. S. 2010, The Astrophysical Journal, 712, 511

Cortes, C., & Vapnik, V. 1995, Machine learning, 20, 273

Cox, D. R. 1958, Journal of the Royal Statistical Society. Series B (Methodological), 20, 215

Debosscher, J., Sarro, L., Aerts, C., Cuypers, J., Vandenbussche, B., Garrido, R., & Solano, E. 2007, Astronomy and Astrophysics, 475, 1159

Efron, B. 1979, The annals of Statistics, 7, 1

Efron, B., & Tibshirani, R. J. 1994, An introduction to the bootstrap (CRC press)

Eyer, L., & Blake, C. 2005, Monthly Notices of the Royal Astronomical Society, 358, 30

Faraway, J., Mahabal, A., Sun, J., Wang, X., Zhang, L., et al. 2014, arXiv preprint arXiv:1401.3211

Freire, A. L., Barreto, G. A., Veloso, M., & Varela, A. T. 2009, in Robotics Symposium (LARS), 2009 6th Latin American (IEEE), 1–6

Grenander, U. 1981, Abstract inference (Wiley New York)

Jiawei, H., & Kamber, M. 2001, San Francisco, CA, itd: Morgan Kaufmann, 5

Kim, D.-W., Protopapas, P., Alcock, C., Byun, Y.-I., & Bianco, F. B. 2009, Monthly Notices of the Royal Astronomical Society, 397, 558

Kim, D.-W., Protopapas, P., Bailer-Jones, C. A., Byun, Y.-I., Chang, S.-W., Marquette, J.-B., & Shin, M.-S. 2014, Astronomy & Astrophysics, 566, A43

Kim, D.-W., Protopapas, P., Byun, Y.-I., Alcock, C., Khardon, R., & Trichas, M. 2011a, arXiv preprint arXiv:1101.3316

—. 2011b, The Astrophysical Journal, 735, 68

Kirk, P. D., & Stumpf, M. P. 2009, Bioinformatics, 25, 1300

Kreiss, J.-P., & Franke, J. 1992, Journal of Time Series Analysis, 13, 297

Kreiss, J.-P., Neumann, M. H., & Yao, Q. 1998

Kunsch, H. R. 1989, The Annals of Statistics, 17, 1217

Lichman, M. 2013, UCI Machine Learning Repository

Mackenzie, C., Pichara, K., & Protopapas, P. 2016, The Astrophysical Journal, 820, 138

Nun, I., Pichara, K., Protopapas, P., & Kim, D.-W. 2014, The Astrophysical Journal, 793, 23

Nun, I., Protopapas, P., Sim, B., Zhu, M., Dave, R., Castro, N., & Pichara, K. 2015, arXiv preprint arXiv:1506.00010

Paparoditis, E., & Politis, D. N. 2000, Annals of the Institute of Statistical Mathematics, 52, 139

Pichara, K., & Protopapas, P. 2013, The Astrophysical Journal, 777, 83

Pichara, K., Protopapas, P., Kim, D., Marquette, J., & Tisserand, P. 2012, Monthly Notices of the Royal Academy Society, 18, 1

Pichara, K., Protopapas, P., & Len, D. 2016, The Astrophysical Journal, 819, 18

Quinlan, J. R. 1986, Machine learning, 1, 81

Quinlan, J. R. 1996, in Proceedings of the Thirteenth National Conference on Artificial Intelligence - Volume 1 (AAAI Press), 725–730

Rasmussen, C. E., & Williams, C. K. I. 2005, Gaussian Processes for Machine Learning (Adaptive Computation and Machine Learning) (The MIT Press)

Richards, J. W., Starr, D. L., Butler, N. R., Bloom, J. S., Brewer, J. M., Crellin-Quick, A., Higgins, J., Kennedy, R., & Rischard, M. 2011, The Astrophysical Journal, 733, 10

Street, W. N., Wolberg, W. H., & Mangasarian, O. L. 1993, in IS&T/SPIE's Symposium on Electronic Imaging: Science and Technology, Vol. 1905 (International Society for Optics and Photonics), 861–870

Udalski, A., Soszynski, I., Szymanski, M., Kubiak, M., Pietrzynski, G., Wyrzykowski, L., Szewczyk, O., Ulaczyk, K., & Poleski, R. 2008, arXiv preprint arXiv:0807.3889

Wachman, G., Khardon, R., Protopapas, P., & Alcock, C. R. 2009, in Machine Learning and Knowledge Discovery in Databases (Springer), 489–505

Wang, P., Khardon, R., & Protopapas, P. 2010, Lecture Notes in Computer Science, 6323, 418